# The ESO's Extremely Large Telescope Working Groups

Paolo Padovani[1]
Michele Cirasuolo[1]
Remco van der Burg[1]
Faustine Cantalloube[2]
Elizabeth George[1]
Markus Kasper[1]
Kieran Leschinski[3]
Carlos Martins[4]
Julien Milli[5]
Sabine Möhler[1]
Mark Neeser[1]
Benoît Neichel[2]
Angel Otarola[1]
Rubén Sánchez-Janssen[6]
Benoît Serra[1]
Alain Smette[1]
Elena Valenti[1]
Christophe Verinaud[1]
Joël Vernet[1]
Olivier Absil[7]
Guido Agapito[8]
Morten Andersen[1]
Carmelo Arcidiacono[9]
Matej Arko[10]
Pierre Baudoz[11]
Olivier Beltramo-Martin[12,2]
Enrico Biancalani[13]
Thomas Bierwirth[1]
Leonard Burtscher[13]
Giulia Carlà[8]
Julio A. Castro-Almazán[14]
Anne-Laure Cheffot[8]
Lodovico Coccato[1]
Carlos Correia[15,2,16]
Romain Fetick[17,2]
Giuliana Fiorentino[18]
Thierry Fusco[17,2]
Begoña García-Lorenzo[14]
Nicola Gentile Fusillo[1]
Oscar Gonzalez[6]
Andrea Grazian[9]
Marco Gullieuszik[9]
Olivier Hainaut[1]
Valentin Ivanov[1]
Melanie Kaasinen[1]
Darshan Kaddad[19,20]
Tomasz Kamiński[21]
Wolfgang Kausch[22]
Florian Kerber[1]
Stefan Kimeswenger[22,23]
Rosita Kokotanekova[24]
Arseniy Kuznetsov[1,17,2]
Alexis Lau[2]
Miska Le Louarn[1]
Frédéric Lemmel[10]
Jochen Liske[25]
Gaspare Lo Curto[1]
David Lucsanyi[26]
Lars Lundin[1]
Stefan Noll[27,28]
Sylvain Oberti[1]
James Osborn[29]
Elena Masciadri[8]
Dinko Milaković[30,31,32]
Michael T. Murphy[33]
Fernando Pedichini[18]
Miguel Pereira Santaella[20]
Roberto Piazzesi[18]
Javier Piqueras López[34]
Cédric Plantet[8]
Thibaut Prod'homme[10]
Norbert Przybilla[22]
Mathieu Puech[35]
Derryck T. Reid[36]
Ansgar Reiners[37]
Rutger Rijnenberg[13]
Myriam Rodrigues[35]
Fabio Rossi[8]
Laurence Routledge[20]
Hans Smit[10]
Mathias Tecza[20]
Niranjan Thatte[20]
Roy van Boekel[38]
Aprajita Verma[20]
Arthur Vigan[2]

1 ESO
2 Aix Marseille University, CNRS, CNES, LAM, Marseille, France
3 Department of Astrophysics, University of Vienna, Austria
4 Centre for Astrophysics of the University of Porto, Portugal
5 Grenoble Alpes University, CNRS, IPAG, France
6 STFC UK Astronomy Technology Centre, Edinburgh, UK
7 STAR Institute, University of Liège, Belgium
8 INAF–Astrophysical Observatory of Arcetri, Firenze, Italy
9 INAF–Astronomical Observatory of Padua, Italy
10 European Space Agency, ESTEC, the Netherlands
11 LESIA, Paris Observatory, CNRS, Meudon, France
12 SpaceAble, France
13 Leiden Observatory, the Netherlands
14 Astrophysics Institute of the Canaries, La Laguna, Spain and Department of Astrophysics, University of La Laguna, Spain
15 Space ODT – Optical Deblurring Technologies, Porto, Portugal
16 Faculty of Engineering, University of Porto, Portugal
17 DOTA, ONERA, Paris Saclay University, Palaiseau, France
18 INAF–Astronomical Observatory of Rome, Italy
19 Space Telescope Science Institute, Baltimore, MD, USA
20 University of Oxford, UK
21 Nicolaus Copernicus Astronomical Center, Toruń, Poland
22 Institute for Astro- and Particle Physics, Leopold Franzens University, Innsbruck, Austria
23 Institute of Astronomy, Catholic University of the North, Antofagasta, Chile
24 Institute of Astronomy and National Astronomical Observatory, Bulgarian Academy of Sciences, Sofia, Bulgaria
25 Hamburg Observatory, Hamburg University, Germany
26 CERN, Switzerland
27 University of Augsburg, Germany
28 German Aerospace Center (DLR), Oberpfaffenhofen, Germany
29 Centre for Advanced Instrumentation, Department of Physics, Durham University, UK
30 Institute for Fundamental Physics of the Universe, Trieste, Italy
31 INAF–Astronomical Observatory of Trieste, Italy
32 National Institute for Nuclear Physics (INFN), Trieste Section, Italy
33 Centre for Astrophysics and Supercomputing, Swinburne University of Technology, Hawthorn, Australia
34 Centre for Astrobiology (CAB), Madrid, Spain
35 GEPI, Paris Observatory, Université PSL, CNRS, Meudon, France
36 Scottish Universities Physics Alliance (SUPA), Institute of Photonics and Quantum Sciences, School of Engineering and Physical Sciences, Heriot-Watt University, Edinburgh, UK
37 Astrophysics und Geophysics Institute, Georg August University, Göttingen, Germany
38 Max Planck Institute for Astronomy, Heidelberg, Germany

**Since 2005 ESO has been working with its community and industry to develop an extremely large optical/infrared telescope. ESO's Extremely Large Telescope, or ELT for short, is a revolutionary ground-based telescope that will have a 39-metre main mirror and will be the largest visible and**

**infrared light telescope in the world. To address specific topics that are needed for the science operations and calibrations of the telescope, thirteen specific working groups were created to coordinate the effort between ESO, the instrument consortia, and the wider community. We describe here the goals of these working groups as well as their achievements so far.**

## Background

In September 2019, ESO's Extremely Large Telescope[1] (ELT) Programme Scientist Michele Cirasuolo, in discussion with several members of the community, as well as the principal investigators of the first-generation ELT instruments (MICADO, MORFEO, HARMONI, and METIS), initiated the formation of a set of working groups (WGs) that had as their main goal the improvement of several critical elements needed by the ELT and its instruments to do transformative science and operate smoothly. These WGs bring together expertise from within ESO, the instrument consortia, and the wider community, with the aim of avoiding redundancy across the consortia, given that many of the issues dealt with are common to all instruments.

At present there are thirteen active WGs (Figure 1), each with its own coordinator(s) and with about 160 contributing members. The overall coordination and the inter-WG deliverables are led by Paolo Padovani (who has replaced Remco van der Burg), and Michele Cirasuolo. The ELT WGs work through various communication channels, including mailing lists, tWiki, Slack, MS Teams, and Zoom. ELT WG meetings have also been held yearly since May 2020.

The ELT WGs are open to the community and volunteers are very welcome. If you are interested in contributing to any of these WGs please contact Paolo Padovani or Michele Cirasuolo[a].

This article introduces the ELT WGs and highlights their main objectives and the results obtained so far.

## Astro-weather (coordinators: Julien Milli and Angel Otarola)

The goal of the Astro-weather WG was twofold. In the first place, it identified the meteorological and atmospheric variables to be monitored, taking into account the requirements of the telescope and each ELT instrument. All these variables were discussed and ranked in three categories according to their priority. In a second step, the WG identified the sensors or technological solutions that could be used to monitor these relevant meteorological and atmospheric variables. This helped to make an estimate of the cost to purchase, and/or design and produce, these sensors. Ultimately, a report was produced summarising all the information gathered by the WG on the requirements for the ELT Astronomical Site Monitor (ASM), including priorities and an estimation of the required budget.

The astro-weather information, to be produced by the various systems comprising the ASM, is essential for the ELT and its instruments to work efficiently, as well as providing important input to other WGs, as shown in Figure 1. Weather data (temperature, relative humidity, wind speed and direction) are relevant to supporting the day and night operations and are therefore considered high-priority. The same is true for turbulence data (seeing, coherence time, isoplanatic angle, and high-resolution profiles of the surface layer turbulence and outer scale) that are used to predict the image quality, to help rank and schedule the science observations, and also to optimally extract the signal (point spread function [PSF] reconstruction: see below). Monitoring the precipitable water vapour is also considered a high priority, to support observations in the infrared (IR) and provide key observations for telluric line corrections (see below). The ELT's main mode of operation will be Service Mode supported by an adaptive queue scheduling of the science observations, and consequently monitoring of the sky transparency also becomes an important factor, as well as forecasting the weather, atmospheric turbulence, and precipitable water vapour on various timescales of interest.

Figure 1. The diagram shows how the different ELT WGs are closely connected, with the output from any given WG feeding directly into (an)other WG(s).

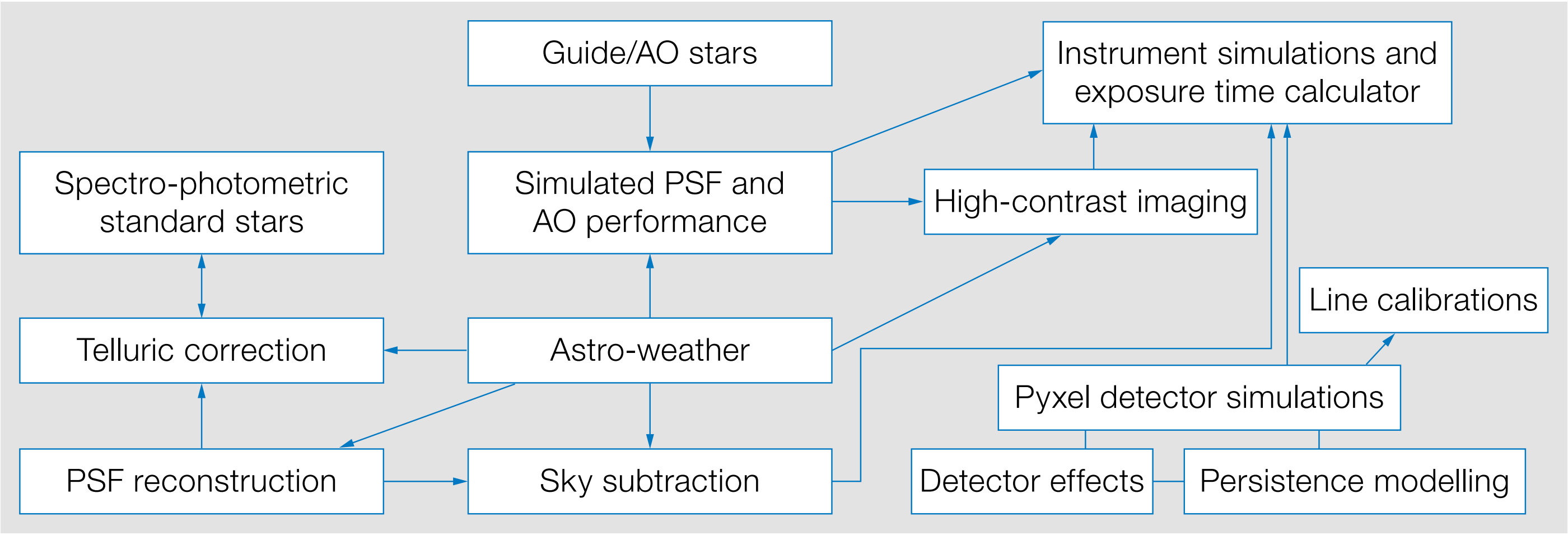

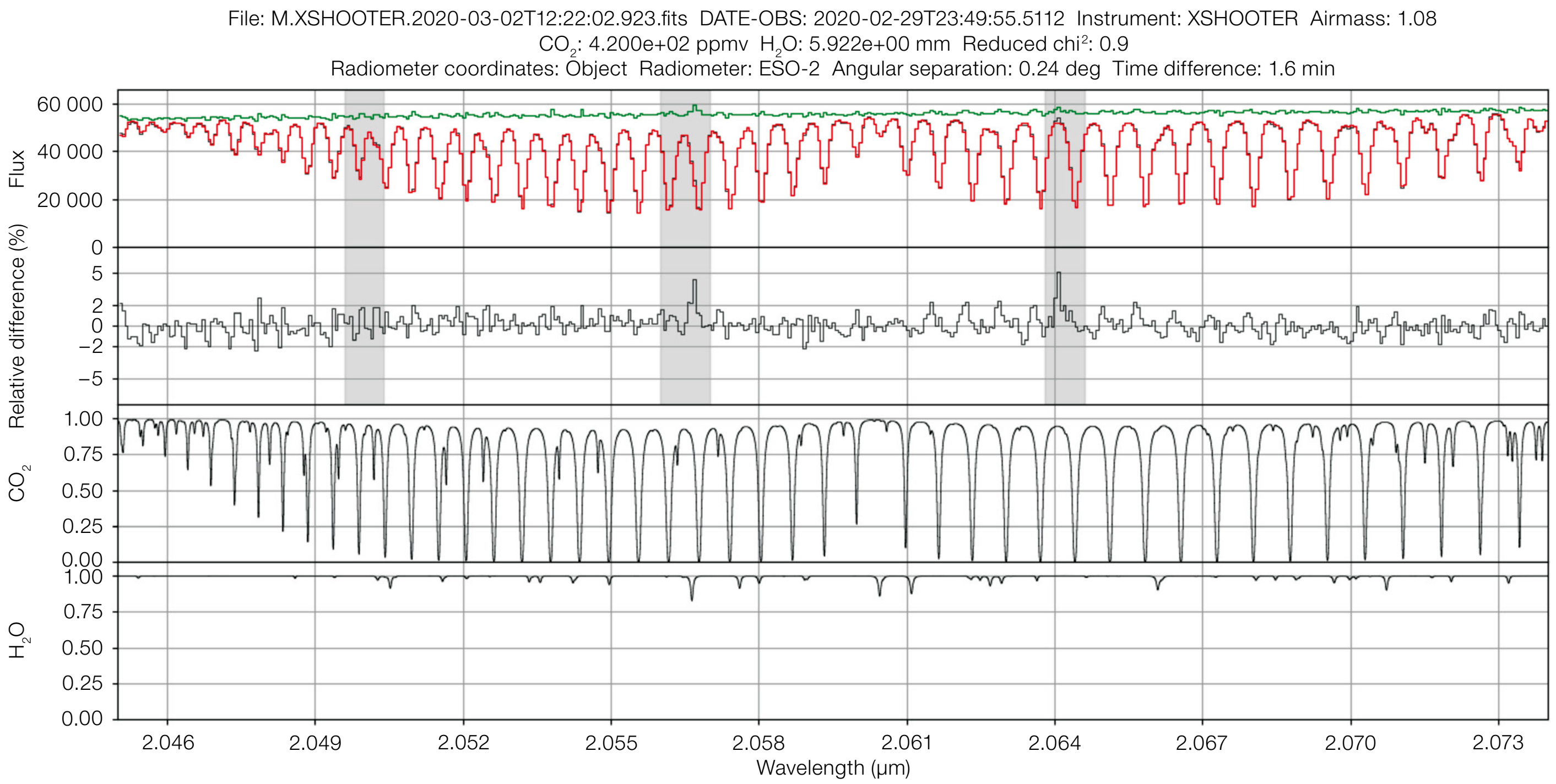

Figure 2. Example of telluric correction of an X-shooter telluric standard spectrum. The top graph shows the X-shooter reduced spectrum retrieved from the science archive in black, the best model obtained by Molecfit in red, and the ratio of the two in green. Data points in the grey area were not used for the fit owing to data quality issues affecting some spectra in a set of several hundred used for the $CO_2$ determination. The bottom two graphs show high-resolution reference (not fitted) spectra for $CO_2$ and $H_2O$. The free parameters of the model include the constants of a 1st-order Chebyshev polynomial for improving the wavelength calibration, the FWHM of a Gaussian profile for the line spread function, and the constants of a 1st-order polynomial representing the continuum. The temperature and humidity profiles (hence, the amount of precipitable water vapour) were obtained at 2020-02-29T23:48:16 with the ESO-2 radiometer, while the relative abundance of $CO_2$ relative to the reference atmospheric model was determined as the mean value of 16 measurements obtained over the month preceding the observations.

## Telluric line correction (coordinator: Alain Smette)

A top-level requirement for the ELT instruments is to minimise nighttime calibrations. However, an IR spectrum displays telluric absorption lines arising from Earth's atmosphere. Their correction usually requires observation of a 'telluric star' (TS) — i.e., a star lacking intrinsic features in the scientifically relevant spectral range — close in time and position to the science target. Over a whole night, the corresponding execution time can easily absorb up to 10% of the science time and changing weather conditions might affect the TS spectrum quality, possibly tarnishing the science observation.

Synthetic telluric absorption spectra have recently provided better correction quality than this empirical method, returning the corresponding overhead time back to science. In particular, Molecfit (Smette et al., 2015; Kausch et al., 2015) is being integrated into instrument pipelines, including those of the ELT instruments (Figure 2). The Molecfit_model routine first adjusts molecular abundances and the parameters of the line spread function (LSF), and possibly corrects for inaccurate wavelength calibration by fitting data in small regions of the science spectra representative of the telluric lines. Then the Molecfit_calctrans routine uses this information to calculate the telluric transmission spectrum over the whole spectral range, so that the Molecfit_correct routine can deliver the corrected spectrum.

The quality of the correction depends on the availability of suitable telluric lines for the fitting process, and hence on the characteristics of the science target spectra. Improvements rely on independently determining the fitted parameters. A microwave radiometer pointing at the target coordinates already provides temperature and humidity profiles (Smette, Kerber & Rose, 2020). Soon, abundances for molecules other than water vapour — which vary on timescales of days to weeks — will be retrieved thanks to regular *twilight* observation of telluric stars. Although the shape of the ELT instruments LSF will be well known, the WG also studies less understood issues, such as the impact of not filling the slit and effects due to the adaptive optics (AO) system.

## Sky subtraction (coordinators: Rubén Sánchez-Janssen and Elena Valenti)

The majority of ELT science cases require observations in the IR, which is notorious for its high sky background. Airglow emission dominates at wavelengths below ~ 2 mm and thermal emission above that, and they must be removed, often down to a few per cent. This is challenging because the airglow lines remain insufficiently characterised — many faint molecular transitions are not yet catalogued — and their high-frequency temporal and spatial variations are poorly constrained. Moreover, we still know very little about the sky continuum emission in the near-IR, except that it is several orders of magnitude brighter than the faint astronomical sources the ELT will be targeting (Oliva et al., 2015).

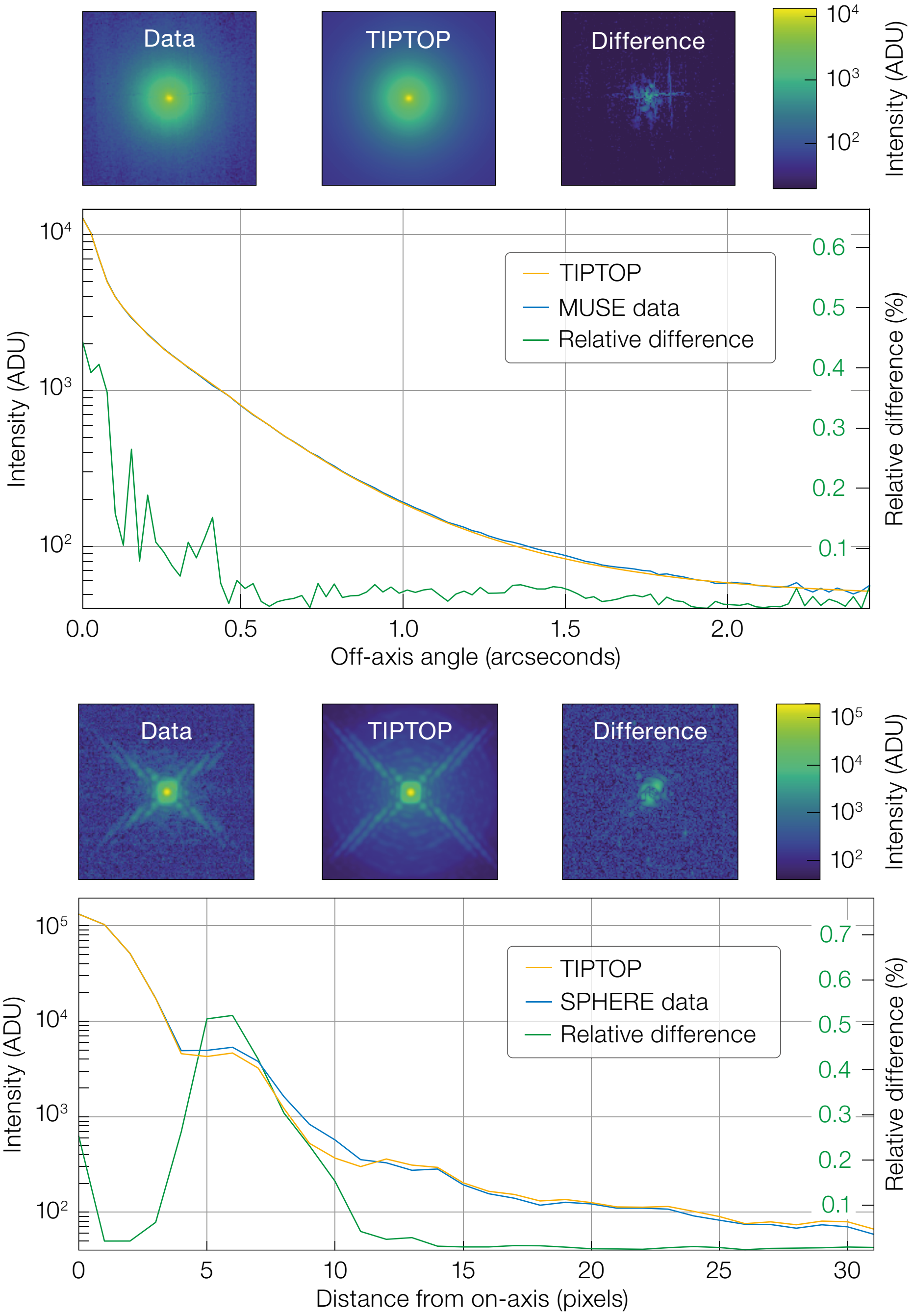

**Figure 3.** Example of TIPTOP PSF fitting for MUSE-NFM (top) and SPHERE (bottom). TIPTOP is used to fit the actual observations, demonstrating that the model is very accurately reproducing the AO PSFs. In operation, TIPTOP will use parameters estimated by the AO and telescope systems to provide an online PSF estimation.

The Sky subtraction WG addresses these issues with the goal of ensuring that ELT instruments deliver sky subtraction to within a few per cent. To this end, it adopts a two-stage approach:

1. Create a precise empirical model of airglow and continuum emission in the near-IR. We are carrying out a comprehensive screening of faint sky lines through a dedicated VLT/CRIRES programme. Additionally, we have embarked on a study of their temporal and spatial variability with archival VLT/X-shooter and Gran Telescopio Canarias/EMIR spectra.

2. We are working on creating and testing sky-subtraction strategies, combining three distinct methodologies: 1) techniques based on a physical model of the sky emission (Noll et al., 2014); 2) probabilistic algorithms based on the statistical properties of the sky signal (Soto et al., 2016); and 3) optimal on-sky observing strategies (Yang et al., 2013). The first two approaches will benefit from the ongoing development of algorithms to characterise the instrumental LSF (Kakkad et al., 2020).

## Spectro-photometric standards (coordinator: Sabine Möhler)

This WG was created to ensure that suitable spectro-photometric standard stars are available by the time the first-light ELT instruments, covering the 0.45 μm to 13 μm wavelength range at various spectral and spatial resolutions, start operating. The spectro-photometric standard stars will be used solely to determine the instrumental response and not for telluric correction (see above). Because the instrumental response is expected to change slowly about 6–10 standard stars per instrument are sufficient, as long as they are evenly distributed in right ascension across the sky and at suitable declinations to avoid observations at large airmass.

For each star reliable reference data must be available across the required wavelength range at the defined brightness intervals. We first investigated whether existing standard star catalogues may be used. It turned out that HARMONI may use the same flux-standard stars as X-shooter and that the standard star catalogues of CRIRES and VISIR are suitable for METIS.

For MICADO, however, new spectrophotometric standard stars need to be defined because the existing ones are too bright. Candidate white dwarf stars have been identified and observed with X-shooter. The analysis is ongoing. The flux-calibrated X-shooter spectra will be fitted with white dwarf model spectra and the best fitting ones will be used as noise-free reference data for the new flux-standard stars.

## Guide/AO stars (coordinators: Paolo Padovani and Giacomo Beccari[b])

Crucial to the operation of the ELT is the availability of stars in the field of view, both for the telescope and the instruments, with the necessary astrometric precision and brightness for telescope acquisition, wavefront control, and AO. The telescope will need up to three natural guide stars (NGS), with information on their optical magnitudes (in the *R* band or *G* band) and good astrometric precision. This can be achieved by using the Gaia stellar catalogue[2], which by the time the ELT operates will have reached its end-of-mission final data release, with accurate parallaxes and proper motions. Depending on the AO mode, the instruments may use NGS in the optical (for example, single-conjugate adaptive optics [SCAO]) or near-IR (laser tomography adaptive optics [LTAO] and multi conjugate adaptive optics [MCAO]).

The work performed so far by this WG includes: 1) the exploration of options to estimate *H*-band magnitudes for stars based on their optical colours and Gaia *G*-band magnitude, since all-sky, near-IR catalogues are not presently available; 2) the determination of the fraction of binaries with close separation (≤ 1 arcsecond and similar brightness (within 3 magnitudes), since these are 'problematic' as ELT guide stars. We find that ~ 20% of stars that could be selected for wavefront sensing are expected to have a companion that could potentially hamper ELT operations; and 3) the drastic reduction of this contamination by using a convolutional neural network approach. The WG at present can produce a guide-star catalogue around user-specified International Celestial Reference System (ICRS) coordinates based on Gaia EDR3, with near-IR fluxes derived from VISTA and UKIRT surveys, and *J*- and *H*-band predictions based on optical data.

## Simulated PSF and AO performance (coordinator: Benoît Neichel)

Since almost all ELT observations will be AO-assisted, the ESO community exposed to AO-corrected data will increase significantly and many future ELT users might not be AO experts. To assist the ESO community in preparing their AO observations, a fast algorithm — called TIPTOP (Neichel et al., 2021) — has been developed, which produces the expected AO PSF for any of the existing AO observing modes (SCAO, LTAO, MCAO, Ground Layer Adaptive Optics [GLAO]) and any atmospheric conditions. Called from a simple application programming interface, TIPTOP is fast enough (a few seconds per PSF) that users can predict the performance of as many configurations as needed, at any sampling, position in the field and wavelength. Moreover, TIPTOP will guide the user to select the best guide star constellation and it will be interfaced with the instrument simulator (see below) to predict the final signal-to-noise ratio (SNR) expected for the target. TIPTOP will also serve for queue scheduling and quality-control, and will provide a first PSF estimation associated with each science observation block. This last step could be seen as a first approach to PSF reconstruction (see below) and may be good enough for some science cases. In preparation for the ELT, TIPTOP will be deployed and tested on various VLT instruments, including ERIS, MUSE, CRIRES, SPHERE, and eventually MAVIS. The WG is currently working towards the fine tuning of the algorithm vs. on-sky observations and first results are very encouraging (Figure 3). TIPTOP has recently been installed as a 'level3 micro-service' on a dedicated ESO machine and is available for beta-testing[3]. Readers are encouraged to test TIPTOP and send feedback.

## PSF reconstruction (coordinator: Joël Vernet)

All currently foreseen ELT instruments will benefit from at least one flavour of AO correction. Developing tools to estimate the highly complex and varying PSFs produced by these systems is therefore crucial to enabling solid measurements of astrophysical quantities (photometric, astrometric, morphological etc.). This need for reconstructed PSFs is further exacerbated by the limited field of view of the ELT instruments and the lack of suitable isolated point sources to estimate the PSF from the science data.

As a starting point, a WG subgroup focused on establishing the state of the art, putting together an overview of the PSF reconstruction approaches currently being explored by various research groups, estimating their respective performance, their range of applicability, and the input and assumptions they rely on. A comprehensive report led by Olivier Beltramo-Martin and others[4] was produced and is available on the WG wiki page.

The most accurate PSF reconstruction (PSFR) algorithms depend heavily on telemetry data produced by the AO systems, such as wavefront sensor data, measured slopes, control matrices or deformable mirror commands at frame rates reaching hundreds of Hz. While extremely data intensive, these methods hold the best potential for reaching percent-level accuracies and there is a clear consensus among WG members that these most promising approaches should be enabled at the ELT. The strategies for AO telemetry data production proposed by the HARMONI, MICADO, and MORFEO consortia were compared and ways to optimise data rates to stay within the practical archiving limit of 10 TB per night were explored (for example, time averaging, pre-processing, compression). Synergies with the Opticon-RadioNet Pilot Joint-Activity JA3.3.2 for virtual access to AO telemetry and development of data storage and exchange standards are also being discussed.

Further topics the WG will focus on include strategies to calibrate non-common path aberrations and the evaluation of PSFR algorithms on current 8-metre-class AO facilities.

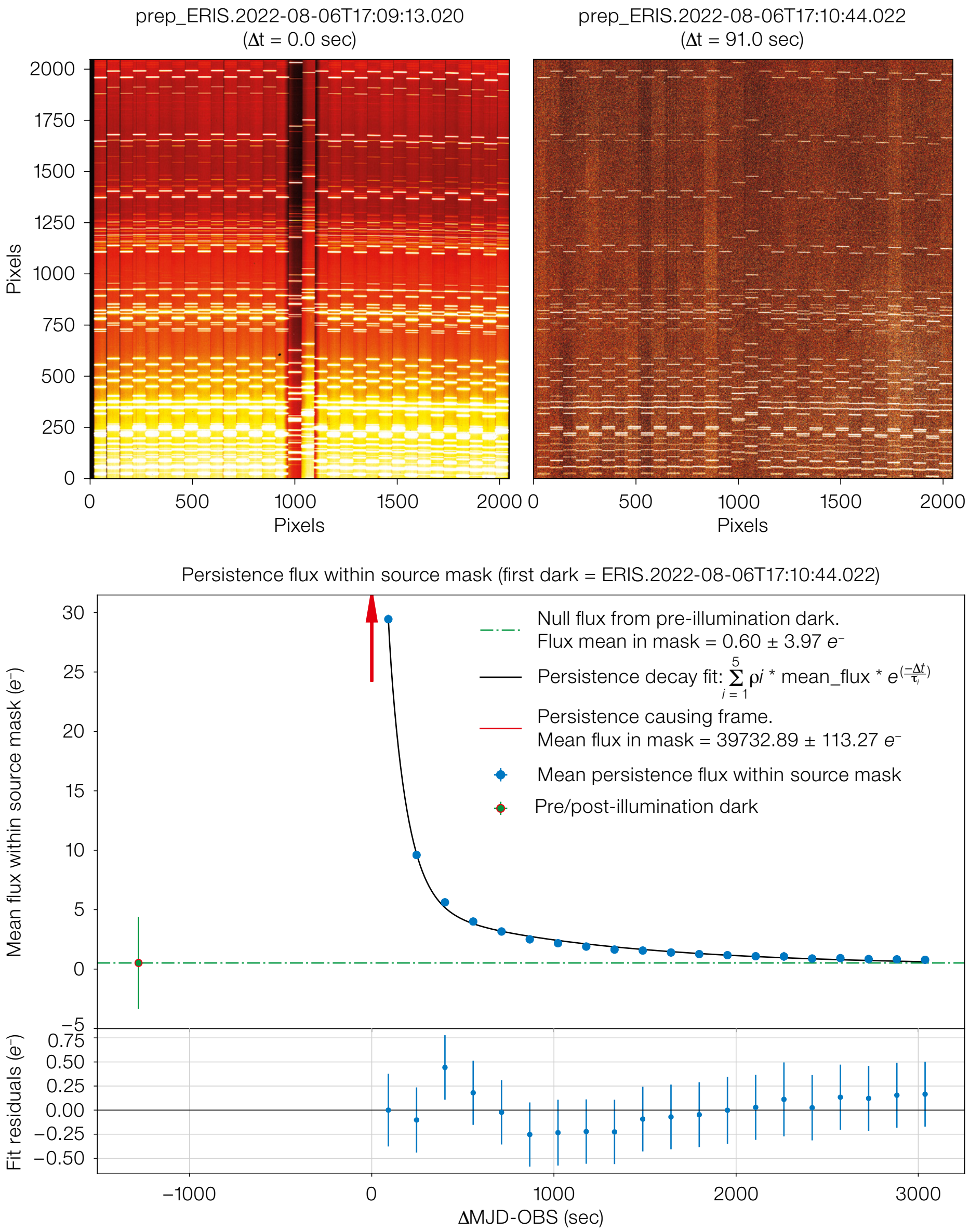

Figure 4. Top Left Panel: An ERIS/SPIFFIER arc lamp calibration, intentionally overexposed to create persistence in a series of dark exposures taken immediately afterwards (peak flux levels are ~ 60 ke⁻).

Top Right Panel: The first 150-second dark following the arc lamp exposure. This dark frame has been dark-corrected by a clean master dark unaffected by persistence. The peak persistence flux levels are ~ 400 e⁻.

Bottom Panel: The exponential decay of the persistence as measured from a series of 20 long darks following the arc lamp exposure. Approximately 2000 seconds after the arc lamp the persistence signal has decayed to zero.

## High-contrast imaging (coordinators: Faustine Cantalloube, Markus Kasper, and Christophe Verinaud)

The three first-light instruments for the ELT (HARMONI, METIS and MICADO), and potentially two of the next-generation ones (ANDES and PCS), include a high-contrast imaging (HCI) mode. This mode is a combination of a high-performance AO correction and an advanced coronagraphic technique that suppresses the starlight diffracted by the telescope aperture before it reaches the science camera. A high-contrast image has a high dynamic range (1:10 000), making it extremely sensitive to wavefront residuals. These residuals might come from uncorrected atmospheric turbulence, inherent limitations of the AO and/or coronagraph, the telescope structure or optical aberrations that are not corrected upstream by the AO. Usually, image processing tailored to HCI data is applied to remove most of these residuals and reach a contrast of 1:1 000 000 or higher. This allows, for example, the detection of mature giant gaseous planets orbiting at a few tenths of an au around low-mass stars, and potentially the circumstellar disc in which they form and evolve.

In this context, the goals of the recently established HCI WG are to: 1) share information, tools, and necessary inputs; 2) report the main limitations foreseen and specific requirements needed for the operations at the ELT; and 3) provide a realistic contrast budget to guide future HCI observations. The three main deliverables are: 1) to implement quick HCI simulations into the ELT simulator; 2) to build the exposure time calculator for HCI modes, with the corresponding observing guidelines; and 3) to prepare for the observation and data exploitation by developing diagnosis and performance tools. This work will benefit common calibration strategies, optimal use of metadata, and definition of scheduling constraints. Therefore, this WG is in close relation to the Astro-weather, AO performance and Instrument simulations and exposure time calculator WGs.

## Line calibrations (coordinator: Carlos Martins)

This WG stems from the strict requirements on the precision, accuracy and stability of cutting-edge astrophysical tests of fundamental physics. One example is that currently useful transitions for measurements of the fine-structure constant need a laboratory wavelength precision of 20 m s$^{-1}$ or better. For the ELT this becomes 4 m s$^{-1}$, implying that many of them require improved laboratory measurements. Such improvements are difficult with current techniques: the natural alternative is to use laser frequency combs (LFCs).

In 2021 the WG made recommendations on ANDES calibration strategies (Martins et al., 2021). In summary: 1) ANDES must have a means to verify wavelength calibration stability requirements, including non-common path errors; 2) monitor novel space and drone-based calibration systems for astronomical telescopes; 3) redundancy in ANDES calibration systems is essential — ANDES should include an iodine cell ($I_2$ cell); 4) one must study whether classical extraction schemes are sufficient for ANDES — delivering precise and accurate uncertainties must be a top-level requirement for the data reduction software; and 5) the wavelength range of any calibrator systems must fully cover the instrument wavelength range.

The WG also recommended that an $I_2$ cell be installed on ESPRESSO, for a twilight calibration programme observing bright, fast-rotating stars to explore the measurement of non-common path and detector effects, how they can be tracked with time, and how accurately they can be removed. A Use Case Proposal was submitted to ESO in November 2021, and the corresponding Change Request is in progress.

The experiment would have two steps: a commissioning run (for two weeks) and a monitoring campaign (every two weeks for one year and around major events). The Big Questions Institute in Sydney has funded the hardware. The cell procurement and calibration are ongoing, and it will be shipped to Paranal when ready so that the experiment can start, provided the ESPRESSO LFC is operational.

### Instrument simulations and exposure time calculator (coordinator: Kieran Leschinski)

The sheer size and complexity of the ELT and its instruments that mandates the use of AO systems in order to observe at the diffraction limit, and the novel science that the ELT will deliver with unprecedented spatial resolution, all require a more advanced instrument simulator to complement the traditional exposure time calculator.

The observation simulator WG has been tasked with creating requirements for a micro-service for the ESO observation preparation environment which can return realistic simulated observations in a timely manner. The working title of this micro-service is ELVIS, the ELT Virtual Instrument Simulator. It will allow users to create 1st-order simulations of their proposed observation, as well as providing SNR estimates in the output format of the chosen observing template. ELVIS will not be created from scratch. The instrument consortia have already invested time and effort into developing instrument data simulators. ELVIS should re-use as much of the existing data and code bases as possible, while taking advantage of the future micro-services developed by other WGs, for example TIPTOP, Pyxel, Skycalc, etc.

Major recent results from the work package include: 1) converging on several key elements of the project scope and its deliverables; 2) the decision to recommend using the ScopeSim generic instrument data simulation ecosystem (Leschinski et al., 2020) as the backbone of the micro-service; 3) definitions of the interfaces with three of the primary external micro-services.

Recent benchmark tests have shown that there should be no major hurdles with implementing the ELVIS micro-service. Readers who do not wish to wait for ELVIS may start experimenting with ELT observations by installing the ScopeSim package in their local python environment.

### Detector effects (coordinator: Elizabeth George). PyXel detector simulations (coordinator: Benoît Serra)

These two closely linked WGs deal with detector performance: detector effects and advanced detector simulations. They are complementary and feed into each other.

The goals of the Detector effects WG are two-fold: 1) to gather knowledge from detector engineers, who through testing in the lab can characterise detector effects that may impact the science, for example, non-linear effects (low count rates), electronic cross talk, persistence, noise, glow; and 2) have scientists analyse the impact on their science resulting from various detector effects.

The main deliverable is a list of common detector effects that can be used as inputs into the Instrument simulations WG (see above). Additionally, the WG provides input to detector groups on a detector characterisation plan (based on detector effects that may impact the science) for each instrument, which includes delivery of standard data products for each detector that can be used to quantify various detector effects.

The advanced (Pyxel) detector simulations WG has the main deliverables of creating simulated detector readouts including all of the detector effects that can be used by various instrument simulators to quantify the impact of detector effects on science data and developing pipeline algorithms to account for these effects.

The two WGs have made good progress towards these goals in the last few years, particularly with the H4RG and H2RG detectors that will be used in the three first-light ELT instruments, HARMONI, MICADO, and METIS.

The Detector Group at ESO has developed standardised characterisation procedures and data products for all the relevant detector effects in the H4RG detectors in MOONS, which will be extended and applied to the detectors for the ELT instruments. This characterisation procedure has been submitted as part of the final design review data packs for MICADO and HARMONI.

Within ESO's collaboration with the European Space Agency (ESA), we have been developing the open-source Pyxel[5] detector simulation framework (Arko et al., 2022), which allows full simulation of detectors and the possibility of implementing any model the user desires. Together with the Pyxel developers at ESA, we held a Detector Modelling workshop in June 2021, which brought together a community of detector engineers and scientists to discuss everything from characterising detectors to developing detector simulators (George et al., 2021). Finally, this year Pyxel has been presented at several conferences (SPIE, EIROforum, SDW2022 and next will be the CMOS workshop at ESA) with several demonstrations of its capabilities. It is now possible to create simulated exposures with H2RG or H4RG including a wide variety of important detector effects using Pyxel and to calibrate some of those models using our laboratory test data.

## Persistence modelling (coordinator: Mark Neeser)

Persistence is the effect whereby a remnant signal from an exposure is imprinted on subsequent images. This effect has long been known to affect HgCdTe near-IR detectors and, if severe, the persistence artefacts can last from hours to several days, negatively affecting the quality of subsequent observations. An example of persistence, intentionally caused within calibration data obtained during ERIS commissioning, is shown in Figure 4. It is hypothesised that persistence is caused by defects and/or impurities within the HgCdTe strata of near-IR detectors. These defects provide sites where light-induced charges are trapped. These trapped charges are generally not released during the detector read but instead randomly decay during subsequent exposures and thereby mimic newly received photo-charge (Smith et al., 2008; Leisenring et al., 2016; Tulloch et al., 2019).

The goal of the Persistence WG is to develop an algorithm and observing strategy for limiting and correcting persistence effects in science data (see Neeser, 2021 for a detailed description). We intend to obtain a deep understanding of how persistence behaves in each new ESO IR detector. This will be done in the laboratory by the ESO Detector Group following a well-defined series of tests. Specifically, this will provide us with a map of the maximum number of persistence traps available in each pixel, the fraction of incident photons that can be converted to persistence traps, and a table of the time constants used to characterise the detector and the relative contribution that each makes to the distribution of persistence traps. A method has been developed to reliably and automatically create persistence maps that can be used to correct any given science exposure.

Using the characterisation data and parameters for each near-IR detector, a model for persistence is used to track the accumulation and decay of persistence traps affecting any input science exposure. These traps are tracked through a series of exposures taken prior to correction of the science frame and a cumulative persistence map is computed for the science frame. Since the data analysed for persistence can be proprietary, this analysis must be done by the ESO Quality Control Group in Vitacura.

The goal is to compute a persistence map for each science exposure and to ingest it into the ESO archive as an associated calibration frame. Since persistence is a rare event, we expect that most maps will contain no significant persistence signal. A blind correction of each science frame for persistence would, therefore, only add noise to the image. Because of this, the subtraction of a persistence map will have to be left to the user. Refining the strategy for this has been helped by the lessons learned during ERIS/SPIFFIER and ERIS/NIX commissioning.

### Acknowledgements

Many thanks to Remco van der Burg who was in charge of the overall coordination of the ELT WGs until February 2022. CJM acknowledges FCT and POCH/FSE (EC) support through Investigador FCT Contract 2021.01214.CEECIND/CP1658/CT0001.

### Links

[1] ELT website: https://elt.eso.org
[2] GAIA catalogue: https://gea.esac.esa.int/archive/
[3] TIPTOP: https://tiptop.readthedocs.io/en/dev/
[4] Subgroup report on PSFR algorithms: https://eso.org/wiki/pub/ELTScience/PSF_reconstruction/ESO_WG_-_PSFR_Algorithms_BeltramoMartin.pdf
[5] Pyxel simulation framework: https://esa.gitlab.io/pyxel/

### Notes

[a] Paolo Padovani: ppadovan@eso.org; Michele Cirasuolo: mciras@eso.org

[b] Until February 2022 the Guide/AO stars WG was lead by Remco van der Burg.